\documentclass[reprint,aps,prl,superscriptaddress]{revtex4-2}
\usepackage{graphicx}
\usepackage{dcolumn}% Align table columns on decimal point
\usepackage{bm}% bold math
\usepackage[usenames]{color}
\usepackage{xcolor}
\usepackage{tabularx}
\usepackage{booktabs}
\usepackage{hyperref}
\usepackage{siunitx}
\usepackage[utf8]{inputenc}
\usepackage[english]{babel}
\hypersetup{colorlinks,linkcolor=red,citecolor=blue,urlcolor=blue,final}
\usepackage{amsmath, amsthm, amscd, amssymb}

\begin{document}
\title{Controlling the sign of optical forces using metaoptics}

\author{Adeel Afridi}
\affiliation{Nanophotonic Systems Laboratory, Department of Mechanical and Process Engineering, ETH Zurich, 8092 Zurich, Switzerland}
\affiliation{Quantum Center, ETH Zurich, 8083 Zurich, Switzerland}

\author{Bruno Melo}
\affiliation{Nanophotonic Systems Laboratory, Department of Mechanical and Process Engineering, ETH Zurich, 8092 Zurich, Switzerland}
\affiliation{Quantum Center, ETH Zurich, 8083 Zurich, Switzerland}

\author{Nadine Meyer}
\email{nmeyer@ethz.ch}
\affiliation{Nanophotonic Systems Laboratory, Department of Mechanical and Process Engineering, ETH Zurich, 8092 Zurich, Switzerland}
\affiliation{Quantum Center, ETH Zurich, 8083 Zurich, Switzerland}

\author{Romain Quidant}
\email{rquidant@ethz.ch}
\affiliation{Nanophotonic Systems Laboratory, Department of Mechanical and Process Engineering, ETH Zurich, 8092 Zurich, Switzerland}
\affiliation{Quantum Center, ETH Zurich, 8083 Zurich, Switzerland}

\date{\today}
%%%%%%%%%%%%%%%%%%%%%%%%%%%%%%%%%%%%%%%%%%%%%%%%%%%%%%%%%%%%%%%%%
\begin{abstract} 

Precise manipulation of small objects using light holds transformative potential across diverse fields. While research in optical trapping and manipulation predominantly relies on the attraction of solid matter to light intensity maxima, here we demonstrate that meta-optics enables a departure from this accepted behavior. Specifically, we present deterministic control over the sign of optical forces exerted on a metasurface integrated on a suspended silicon nanomembrane. By tailoring the geometry of the constituent meta-atoms, we engineer the coherent superposition of their multipolar modes, and consequently, the net optical force experienced by the metasurface within a phase-controlled optical standing wave. In excellent agreement with 3D numerical simulations, we experimentally realize both attractive and repulsive forces on distinct metasurface designs, directly mirroring the behavior of two-level systems interacting with optical fields. This work establishes a versatile platform for the optical control of nanoscale mechanical systems, opening new avenues for both fundamental research and engineering.

\end{abstract}
%%%%%%%%%%%%%%%%%%%%%%%%%%%%%%%%%%%%%%%%%%%%%%%%%%%%%%%%%%%%%%%%%

\maketitle

% Ashkin papers
%Acceleration and trapping of particles by radiation pressure~\cite{ashkin1970acceleration}
% Optical Levitation by Radiation Pressure \cite{ashkin1971optical}
% Observation of a single-beam gradient force optical trap for dielectric particles ~\cite{ashkin1986observation}
%Observation of Resonances in the Radiation Pressure on Dielectric Spheres~\cite{ashkin1977observation}

%First Kerker condition met for crystalline Si at 1064nm for r = 131nm

%resonant change in RPF implies a resonant change in light scattering since optical force arises directly from light scattering~\cite{ashkin1981observation} (2nd paper)

%%%%%%%%%%%%%%%%%%%%%%%%%%%%%%%%%%%%%%%%%%%%%%%%%%%%%%%%%%%%%%%%%%%%%%%

The interaction of light with matter involves the fundamental exchange of photon momentum, resulting in radiation pressure. This phenomenon, predicted by Maxwell's electromagnetic theory, was experimentally confirmed in 1901 by Lebedev~\cite{lebedew1901untersuchungen}, and Nichols and Hull~\cite{nichols1901preliminary}, followed by the observation of photon recoil by Poynting and Barlow~\cite{poynting1910bakerian}. Although radiation pressure under terrestrial conditions is usually too small to be noticed, it plays an important role in the formation of stars~\cite{murray2009disruption} and in the dynamics of spacecrafts in outer space~\cite{anselmo1983orbital,scheeres1999satellite}, where the radiation pressure is the main force next to gravity. Nowadays, radiation pressure is even harnessed on demand, for instance in photonic sails %for the ambitious approach 
to propel ultralight satellites through space~\cite{ilic2019self,michaeli2025direct,spencer2021lightsail} or to control optomechanical systems across a wide range of sizes~\cite{abramovici1992ligo,cohadon1999cooling,aspelmeyer2014cavity}.

However, light-matter interaction is not limited to repulsive forces. Ashkin's seminal work revealed that polarizable objects placed in an inhomogeneous light field experience a force that attracts them towards the highest intensity~\cite{ashkin1986observation, ashkin1971optical,ashkin1970acceleration}. This discovery laid the foundation for the field of optical tweezers, enabling the precise control and manipulation of microscopic particles. Nowadays, optical tweezers are widely used in cell-biology~\cite{zhang2008optical,goel2008harnessing,zhu2020optical}, climate research with aerosols~\cite{burnham2006holographic}, quantum optics and quantum simulations with ultracold atoms and molecules~\cite{grimm2000optical,kaufman2021quantum}, and levitation optomechanics~\cite{gonzalez2021levitodynamics,millen2020optomechanics}, among others. Further experimental exploration revealed surprising effects, such as pulling forces from engineered unfocused light beams, coined tractor beams 
(see discussion in~\cite{chen2011optical, ding2019photonic,shvedov2014long,brzobohaty2013experimental}) or non-reciprocal optical binding~\cite{dholakia2010colloquium,rieser2022tunable}. %, allowing for naturally enhanced optical forces~\cite{kaplan2002optimized,stilgoe2008effect,li2013giant,andres2016optical,lank2018directional}. 

% 2 INTRODUCTION ===========================================================
The optical forces experienced by a specimen exposed to a light intensity gradient are hereby fundamentally governed by its polarizability, offering additional control through its geometrical and structural properties. This was first highlighted in 1977 when Askhin and Diedzic observed enhanced radiation pressure on levitated oil drops from a probe beam spectrally matching their Mie resonances ~\cite{ashkin1977observation}. Similar resonant effects were also studied in plasmonic nanoparticles~\cite{dienerowitz2008optical, arias2003optical,zelenina2006tunable} and nanodiamonds hosting multiple nitrogen-vacancies~\cite{juan2017cooperatively}. %exploiting structural (electronic) resonances. In ~\cite{juan2017cooperatively} nanodiamonds hosting multiple nitrogen-vacancy (NV) centers were shown to exhibit enhanced optical forces at their zero-phonon line. 
More recently, drawing an analogy to two-level atoms~\cite{vetsch2010optical}, we proposed to exploit electromagnetic Mie resonances in high-permittivity meta-atoms for optical trapping at intensity minima~\cite{lepeshov2023levitated}. Furthermore, unconventional switchable optical forces 
were reported in nanoparticles made of temperature-sensitive phase change materials ~\cite{mao2024switchable}.

In this work, we demonstrate full control over optical forces on a metasurface~\cite{chen2016review,yu2011light,kruk2017functional,kuznetsov2024roadmap}. Through precise meta-atom design, we experimentally achieve a controllable reversal of the optical forces, from repulsive to attractive, in excellent agreement with 3D multi-physics simulations. Leveraging solely geometry and intrinsic material properties, our method generates these forces without complex beam engineering, offering new possibilities for scalable light-based manipulation of matter.

\begin{figure}
\centering
\includegraphics[width=0.48\textwidth]{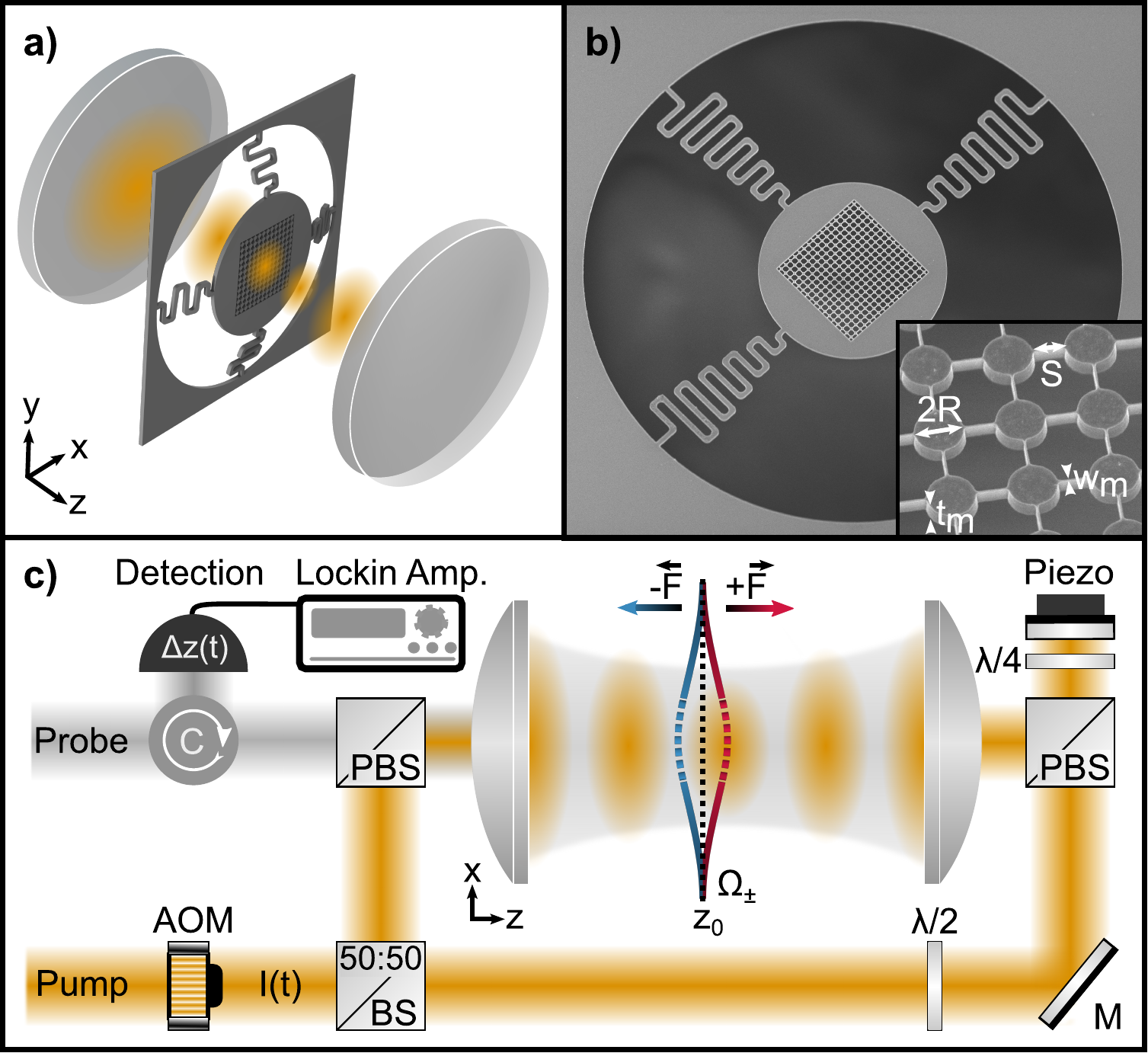}
\caption{\textbf{Experimental configuration.} \textbf{a,}  A silicon membrane, patterned with a metasurface, is suspended within the standing wave created by two phase-stabilized, counterpropagating beams focused by a pair of lenses. Undulated tethers allow the free standing membrane to move along $z$.  \textbf{b,} SEM picture of tethered, patterned membrane of thickness $t_m$ featuring a symmetric disc array with disc radius $R$, separated by connectors of length $S$ and width $w_m$ (see inset). \textbf{c,} The suspended membrane with mechanical eigenfrequency $\Omega_\pm$ is positioned at $z_0$ along the standing wave (yellow). The standing wave phase $\phi_0$ is controlled with a piezo-actuated mirror. The light intensity $I(t)$ is modulated by an acousto-optical modulator (AOM). The force vector $F_+$ ($F_-$) indicates a membrane displacement towards $z > z_0$ ($z < z_0$). The probe beam (gray) is phase modulated by $z(t)$ and used for homodyne detection of the membrane displacement where a lock-in detection extracts both phase $\Theta(\omega)$ and amplitude $\mathcal{A(\omega)}$.  }
\label{fig:1}
\end{figure}

% 3 IDEA BEHIND THE EXPERIMENT ================================================================
\paragraph{\textbf{System description}}
The studied configuration features a free-standing metasurface placed in an optical standing wave, as illustrated in Fig.~\ref{fig:1}a. The metasurface design consists of a %$N\times N$ 
periodic array of identical silicon discs, each with radius $R$, separated by short connectors of length $S$ (see Fig.~\ref{fig:1}b). To minimize mechanical stress and allow free motion, the metasurface is suspended from a frame using undulated tethers~\cite{chou2015fabrication}.  Given that a single optical beam carries  %with 
significant photon momentum along its propagation direction, it cannot generate a pulling force~\cite{chen2011optical}. Hence, we opted for a standing wave field formed by two counter-propagating Gaussian beams. This configuration cancels the scattering force, thereby ensuring that the gradient force dominates.  Within the Gaussian beam approximation, the gradient force experienced by the metasurface is given by $F (z) = \alpha_{\text{eff}} \nabla I(z)$ where $\alpha_{\text{eff}}$ represents the real part of the effective polarizability, and $\nabla I(z)$ the intensity gradient at the metasurface position $z_0$. The effective polarizability  $\alpha_{\text{eff}}$ is primarily determined by the magnetic and electric dipolar, quadrupolar and octupolar modes~\cite{chen2011optical,lepeshov2023levitated} supported by each meta-atom.

Using this configuration, we aim to demonstrate accurate control over both the magnitude and direction of the total force by tuning geometrical parameters to engineer the coherent superposition of Mie multipolar modes ~\cite{chen2011optical}. Specifically, we leverage the dependence of the scattering radiation pattern of silicon discs, and hence the net optical force $F_0$ they experience, on the ratio of their radius $R$ to the laser wavelength $\lambda$~\cite{lepeshov2023levitated}.  By varying disc radii $R$ and separations $S$, we explore a wide range of force amplitudes and directions. Depending on the optical wavelength, each metasurface acts as a high- or low-field seeker attracted to intensity maxima or minima, in full analogy to two-level systems~\cite{vetsch2010optical}.

\paragraph{\textbf{Theoretical model}}
To estimate the total optical force experienced by the freely moving metasurface, we exploit its resonant mechanical mode, which exhibits a frequency-dependent amplitude response to an external driving force~\cite{cuairan2022precision,ricci2019accurate}. 
%To estimate the tailored optomechanical response of the metasurface to an external optical driving force $F(t)$, we 
We model the system as a driven underdamped harmonic oscillator, %that follows the equation of motion:
%
%\begin{equation}\label{eq:EoM}
%    \Ddot{z}(t) + \Gamma \Dot{z}(t) + \Omega^2 z(t)= \frac{F(t)}{m}
%\end{equation}
characterized by the oscillator's mechanical eigenfrequency $\Omega$, its mass $m$ and the amplitude decay rate $\Gamma$ due to clamping losses and gas collisions at atmospheric pressure.
%with $\Omega$ being the oscillator's mechanical eigenfrequency, $m$ its mass, $\Gamma$ the mechanical damping due to clamping losses and gas collisions at atmospheric pressure.
%To d
The optical driving force %$F(t)\propto I(t)$
is generated by modulating the intensity of the standing wave pattern $I(t) = I_0 (1 -\cos(\omega_{\text{dr}}t))/2$, such that  a time-dependent optical force $ F(t)=F_0\: (1 -\cos(\omega_{\text{dr}}t))/2$ is exerted on the membrane. In this context, the optical driving force 
 $F(t)$ prevails over the thermal driving force (see Supplementary Information Sec.~V).
The displacement of the oscillator, $z(t)= \mathcal{A}(\omega) \cos(\omega t + \theta(\omega))$, is characterized by its frequency dependent amplitude
$\mathcal{A}(\omega)\propto F_0$ 
%\begin{equation}\label{eq:amplitude}
%\mathcal{A}(\omega) = \frac{F_0}{2m \sqrt{(\Omega^2 - \omega^2)^2 + (\Gamma \omega)^2}},    
%\end{equation}
%
and its phase $\theta(\omega)$ across the resonance.
%\begin{equation} \label{eq:theta}
%\theta(\omega) = \tan^{-1}\left( \frac{\Gamma \omega}{\Omega^2 - \omega^2} \right).    
%\end{equation}

Eventually, the sign of the intensity gradient at the membrane's equilibrium position along the standing wave dictates the direction of the displacement. While the amplitude and phase relationship between $z(t)$ and $F(t)$ are invariant for high- and low-field seekers, their displacement directions are opposite (see Fig.~\ref{fig:1}c). However, since measurements are referenced to the driving signal of $I(t)$, this results in an effective reversal of the force $F(t)$ and a phase shift in the measured motion $z(t)$ relative to the modulation signal, when comparing high- and low-field seekers at the same position.
%
%The two observed phases are then given by
The measured phase between $I(t)$ and $z(t)$ can thus take on two distinct values, given by:
\begin{equation}\label{eq:Theta}
    \Theta(\omega) =  \theta(\omega) + \phi = \begin{cases}
\theta(\omega) + 0 \: \text{for}  \: F_+\\
\theta(\omega) + \pi\: \text{for} \: F_-
\end{cases}
\end{equation}

where we refer to in-phase (out-of-phase) oscillations as positive (negative) force $F_+$ ($F_-$). Note that, depending on the sign of the intensity slope, both high- and low-field seekers can exhibit positive and negative force behavior. To account for this we define %the phase term
$\phi = |\phi_0 + \phi_s|$. The phase term $\phi_s=+\pi/2$ ($\phi_s=-\pi/2$) corresponds here to a low- (high-) field seeker that is repelled (attracted) by high intensity. This consequently leads to $\phi_0=+\pi/2$ ($\phi_0=-\pi/2$) for the positive (negative) intensity slope of the standing wave pattern. \\
%The phase term $\phi_0=+\pi/2$ ($\phi_0=-\pi/2$) corresponds here to the positive (negative) intensity slope of the standing wave pattern. This consequently leads to $\phi_s = -\pi/2$ ($\phi_s = + \pi/2$) for a high (low) field seeker that is attracted (repelled) by high intensity.\\

% 4 OVERVIEW OF THE EXPERIMENT/PLATFORM =====================================================
\paragraph{\textbf{Experimental implementation}}
%\paragraph{Experimental settings}

The experimental configuration is illustrated in Fig.~\ref{fig:1}c.  Two counter-propagating, equally $y$-polarized beams at a wavelength $\lambda=\SI{1550}{\nano\meter}$ with power $P=\SI{20}{\milli \watt}$ are focused by two lenses of numerical aperture NA=0.4 resulting in a beam waist smaller than the metasurface area. Both beams are phase-stabilized to form an interference pattern along the optical axis $z$ (shown in yellow) where the relative position of intensity maxima and membrane position $z_0$ is controlled with the beams' relative phase $\phi_0$. The optical external driving force is generated by modulating the optical intensity $I(t)$ of this standing wave with an acousto-optic modulator (AOM) at $\omega_\text{dr}$ and amplitude $I_0$. The metasurface, consisting of an array of identical discs, is patterned on %commercially available, 
freestanding crystalline silicon membranes of thickness $t_m=\SI{350}{\nano\meter}$~\cite{afridi2023ultrathin}. The discs in the array are connected by thin nanobeams of width $w_m=\SI{70}{\nano\meter}$. 
%To detect the optical force $F_0$, we utilize the resonant behavior of the freely moving membrane, which exhibits a frequency-dependent amplitude response to an external driving force. 
%To detect the motion of the driven membrane, 
An additional single cross-polarized beam co-propagates along the optical axis (depicted in gray in Fig.~\ref{fig:1}c). The backscattered light of this probe beam is phase modulated by the motion of the membrane $z(t)$ and therefore can be used for optical displacement readout via phase sensitive homodyne detection~\cite{tebbenjohanns2021quantum}. By using a lock-in operation, we detect the membrane motion $z(t)= \mathcal{A}(\omega_\text{dr}) \cos(\omega_\text{dr}t +\theta(\omega_\text{dr}) + \phi) $ with the relative phase $\theta(\omega_\text{dr}) + \phi$ between the membrane motion and the reference signal $\propto \cos(\omega_{\text{dr}}t)$ (see Supplementary Information Sec.~III). \\%, while simultaneously scanning the driving frequency $\omega_{\text{dr}}$ at constant intensity amplitude $I_0$.

% 5 Demonstration of negative and positive forces  (Fig 2) =========================================================
%\paragraph{Demonstration of negative forces}
\begin{figure}
\centering
\includegraphics[width=0.5\textwidth]{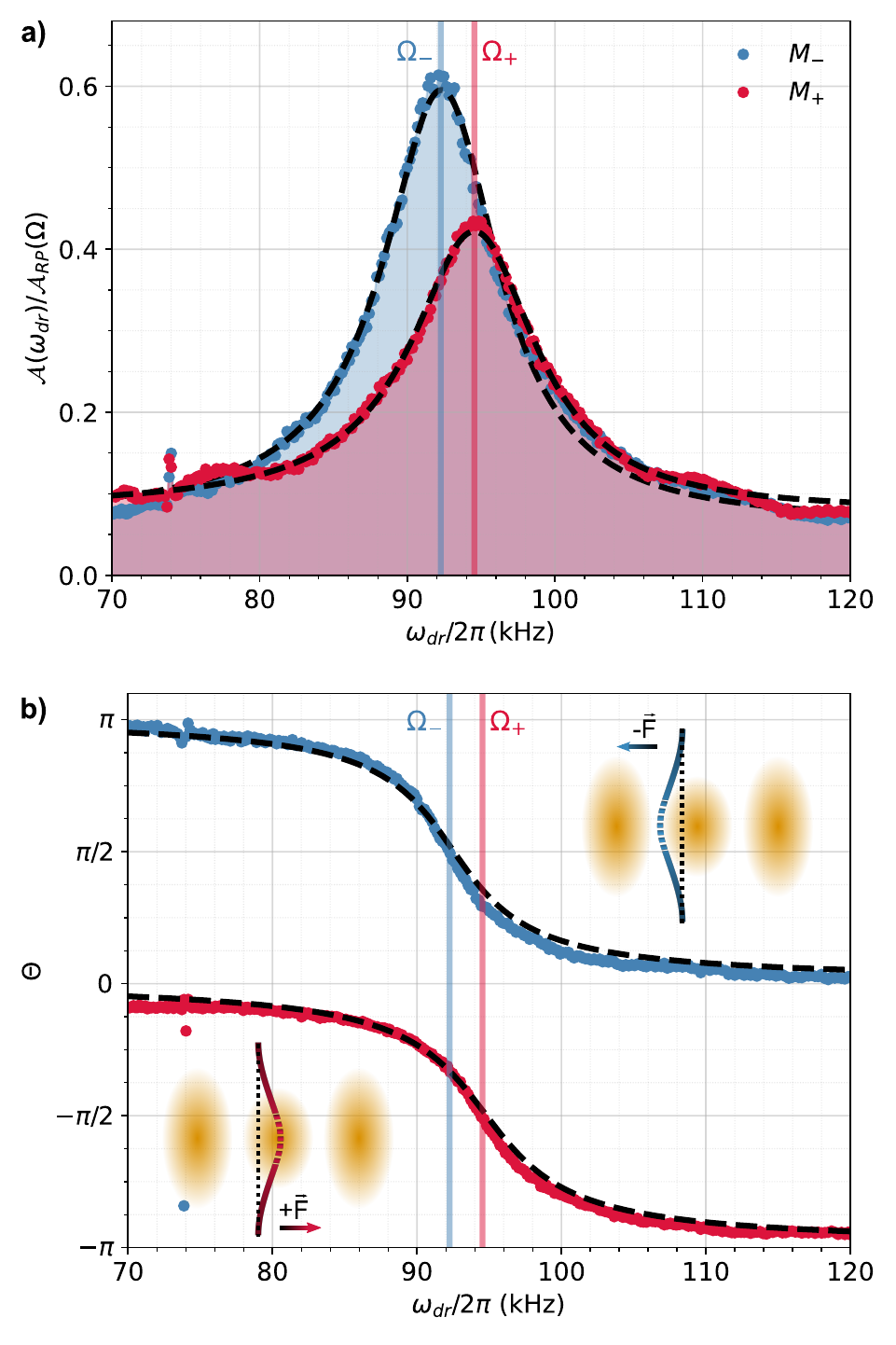}
\caption{\textbf{Metasurfaces with positive and negative force response $F_\pm$.} \textbf{a,} Normalized driven displacement amplitude $\mathcal{A}(\omega_{\text{dr}})/\mathcal{A}_{\text{RP}}(\Omega)$ for membranes $M_+$ (red)  and $M_-$  (blue)  around their respective resonance $(\Omega_{+}, \Omega_{-})/(2\pi) = [94.81,92.55]\SI{}{kHz} $ (solid lines) at atmospheric pressure. \textbf{b,} Driven phase response $\Theta(\omega_{\text{dr}})$ for $M_+$ (red) is in-phase ($|\Theta(\omega_{\text{dr}})|\approx 0$) below the resonance ($\omega < \Omega_+$) and out-of-phase ($|\Theta(\omega_{\text{dr}})|\approx \pi$) above the resonance ($\omega > \Omega_+$), indicating a positive force response $F_+$. The $M_-$ membrane (blue) shows a negative force response $F_-$, demonstrating the opposite phase response ($|\Theta(\omega_{\text{dr}})|\approx \pi$ for $\omega < \Omega_-$ and $|\Theta(\omega_{\text{dr}})|\approx 0$ for $\omega > \Omega_-$).  Dashed lines are fits to the theory and the solid lines highlight the resonance frequency $\Omega_{\pm}$. The insets indicate the membrane motion for $F_\pm$ towards high or low intensities.}
\label{fig:2}
\end{figure}

\begin{table}[h]
   \begin{tabular}{|c|c|c|c|c|c|}
\hline
  & $R$ [nm] & $S$ [nm] & $\Omega_{\pm}/(2\pi)$ [kHz] & $\Gamma_{\pm}/(2\pi)$ [kHz] & $\phi_s$  \\
\hline
\hline
$M_+$ & 345 &530 & 94.81 & 10.32 & $-\pi/2$ \\
\hline
$M_-$ &485 & 430 & 92.55 & 9.54 & $+\pi/2$ \\
\hline
\end{tabular}
\caption{\textbf{Parameters for membrane $M_+$ and $M_-$} used in Fig.~\ref{fig:1} and Fig.~\ref{fig:4} with radius $R$, separation $S$, mechanical eigenfrequency $\Omega_\pm$, damping $\Gamma_\pm$ and $\phi_s$. The parameters $\Omega_\pm$ and $\Gamma_\pm$ are obtained from the fit in Fig.~\ref{fig:1}a.}
 \label{tab:1}
\end{table}

We first focus on two membranes (see Tab.~\ref{tab:1}), $M_+$ and $M_-$, which exhibit high- and low-field seeker behavior, respectively. These membranes are positioned at the rising slope of the intensity field ($\phi_0=\pi/2$). %The specific parameters are $r = 345\,\text{nm}$ and $s = 530\,\text{nm}$ for $M_+$, and $r = 485\,\text{nm}$ and $s = 430\,\text{nm}$ for $M_-$.  

Fig.~\ref{fig:2} depicts the motional response of membranes $M_+$ and $M_-$ under the driving force $F(t)$ with varying $\omega_\text{dr}$ at atmospheric pressure. 
The amplitude $\mathcal{A}_{\pm}(\omega_\text{dr})$ is normalized by  the experimental amplitude response of an unstructured flat membrane $\mathcal{A}_{\text{RP}}(\Omega)$ acting as a mirror. This mirror membrane experiences radiation pressure force from a single beam $F_{\text{RP}} = \frac{P}{c}[2r + a]$~\cite{swartzlander2017radiation}, where $r$ and $a$ are the reflectivity and absorption of the unstructured flat membrane, $P = \SI{40}{\milli\watt}$ the optical power, and $c$ the speed of light (see Supplementary Information Sec.~IV).

As shown in Fig.~\ref{fig:2}a, the driven motion exhibits the typical Lorentzian profile of a mechanical resonance, centered at $\Omega_{\pm}$ (solid lines) and with a linewidth $\Gamma_\pm$ (see Tab.~\ref{tab:1}). %predicted by Eq.~\ref{eq:amplitude}. 
%As depicted in Fig.~\ref{fig:2}a, the driven motion follows the Lorentzian shape   with typical mechanical resonances at $(\Omega_+,\Omega_-) = 2\pi\times [94.81, 92.55]\SI{}{\kilo\hertz}$ (dashed lines) and linewidth $(\Gamma_+,\Gamma_-) = 2\pi\times [10.32, 9.54]\SI{}{\kilo\hertz}$, yielding quality factors of $(Q_+,Q_-)=[9.19,9,7]$. 
Fig.~\ref{fig:2}b displays the phase response $\Theta(\omega_{\text{dr}})$ of the driven membranes $M_+$ and $M_-$. The dashed lines represent fits to the theory (see Supplementary Information Sec.~III). %Eq.~\ref{eq:amplitude}-\ref{eq:theta}. 
We observe that the membrane $M_+$ (red) oscillates in phase with the reference signal at low $\omega_\text{dr}$. Upon exceeding its resonance frequency ($\omega>\Omega_+$) it undergoes the expected $\pi$ phase jump. %from  $\Theta=0$ to  $\Theta=-\pi$ 
indicating that the oscillator response lags behind the driving force. In contrast, $M_-$ oscillates out of phase ($\Theta=\pi$) already for $\omega < \Omega_-$  and changes to an in-phase oscillation with respect to the reference signal ($\Theta=0$) across the resonance.  The different phase responses, in excellent agreement with the theory, demonstrate that the membranes $M_+$ and $M_-$ experience opposite force signs (see insets in Fig.~\ref{fig:2}b). Specifically, $M_+$ acts as a high-field seeker, drawn to the intensity maximum, whereas $M_-$ behaves as a low-field seeker, drawn atypically to the intensity minimum, a scenario difficult to achieve with dielectric structures in the dipole regime~\cite{dago2024stabilizing}.  

% 6 Theory discussion of negative forces   (Fig 3) ==========================================
%\paragraph{Force tunability versus size, force enhancement}
\begin{figure*}
\centering
\includegraphics[width=1\textwidth]{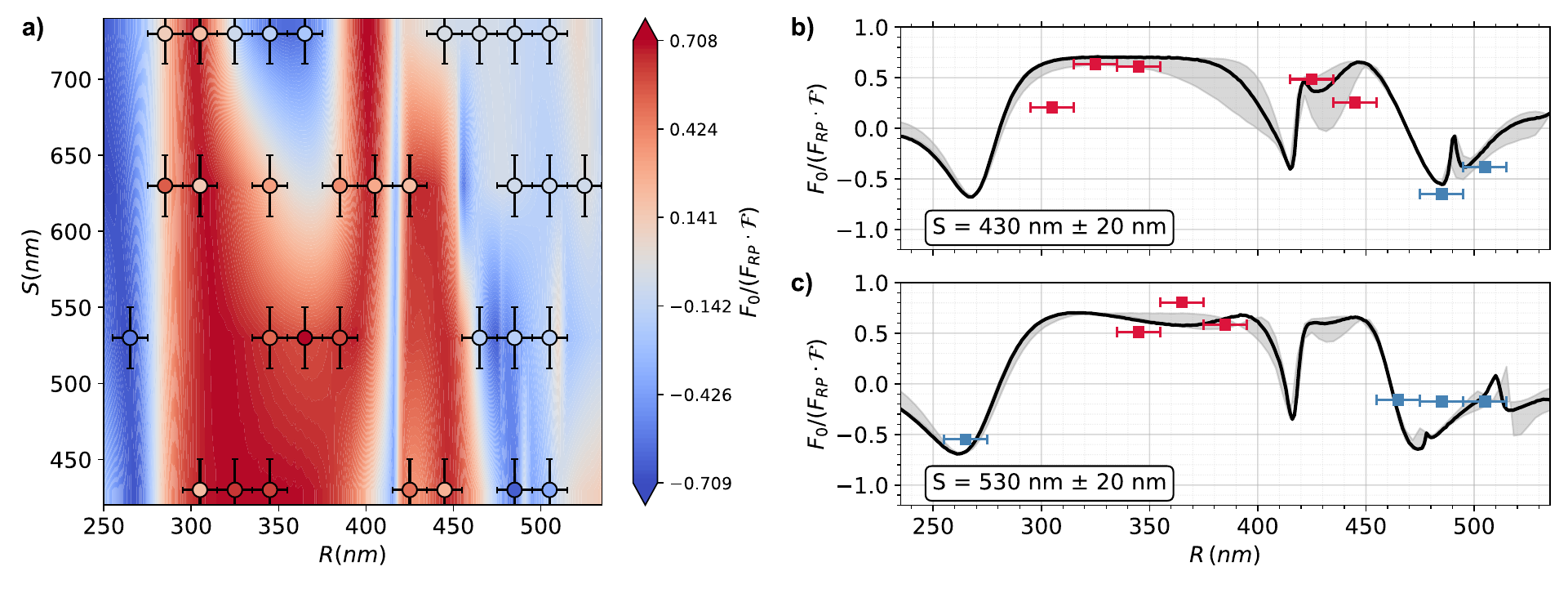}
\caption{\textbf{Optical force dependence on the geometrical parameters of the metasurface.} \textbf{a,} Comparison of simulated (contour) and experimental optical forces (circles) as function of disc radius $R$ and separation $S$.  Positive forces $F_+$ ($>0$, red) and negative forces $F_-$ ($<0$, blue) are achieved by tuning $R$ and $S$. Dependence of the optical force on the radius $R$ at \textbf{b,} $S=\SI{430}{nm}$  and \textbf{c,}  $S=\SI{530}{nm}$.  Experimental results (red and blue squares for positive and negative forces, respectively) show a non-trivial dependence on the radius, consistent with COMSOL simulations (solid lines).
Both experimental data and simulations are normalized to the radiation pressure force of a mirror corrected by the filling factor $\mathcal{F}$ (see Supplementary Information Sec.~IV). Error bars on the experimental data represent uncertainties due to fabrication tolerances $\delta R =\pm \SI{10}{nm}$ and $\delta S =\pm \SI{20}{nm}$. The gray shaded area accounts for the fabrication error in the simulations.}
\label{fig:3}
\end{figure*}
%To investigate the dependence of structural parameters and complete control over the optical force,

To demonstrate complete control over the optical force and its dependence on structural parameters, we fabricate a series of metasurfaces systematically exploring the $(R, S)$ parameters space where the radius $R$ ranged from $250$ to $550\SI{}{\nano\meter}$ and the separation ranged from $430$ to $730\SI{}{\nano\meter}$. For each metasurface, we position the membrane at $\phi_0=\pi/2$ and modulate the intensity at $\omega_{\text{dr}}= \Omega$ for each metasurface. 

In Fig.~\ref{fig:3}a we compare the measured force amplitudes $F_0/(F_{\text{RP}}\mathcal{F}) = [\mathcal{A}_\pm(\Omega_\pm)\Gamma_{\pm}\Omega_{\pm}] / [\mathcal{A}_{\text{RP}}(\Omega) \Gamma_{\text{RP}}\Omega_{\text{RP}}]$ (circles) with simulation. The geometric filling factor $\mathcal{F} = (\pi R^2 + 2S w_m)/(2R + S)^2$ is equal to the mass ratio of the metasurface and mirror membrane. The contourplot displays the simulations (see Supplementary Information Sec.~II). Fig.~\ref{fig:3}b and c show line plots for constant separations $S=\SI{430}{\nano\meter}$ and $\SI{530}{\nano\meter}$ exhibiting the largest negative force $F_-$ and largest positive force $F_+$ with a magnitude comparable to the radiation pressure $F_\text{RP}$. The fabrication uncertainty of $\delta R =\pm \SI{10}{\nano\meter}$ and $\delta S = \pm \SI{20}{\nano\meter}$ is represented by the experimental error bars and the gray-shaded area in the simulation. %, while in the simulation we assume %attribute this  uncertainty to the separation $s$ 
%$\delta S = \pm \SI{20}{\nano\meter}$ combined with $\delta R= \mp \SI{10}{\nano\meter}$ (gray shaded area). 
We observe a broad tunability of the force $F_0$ that is only weakly affected by the separation $S$. However, at smaller radii, the force amplitude tends to be larger with frequent negative values, whereas at midrange radii, the amplitude is predominantly positive. We find experiment and simulation in good agreement, confirming the tunability of optical forces by structural parameters $R$ and $S$.

%To this aim, we normalized the amplitude $\hat{Z}$ by the amplitude of a driven flat membrane $\hat{Z}_{mirror}$, at their respective mechanical eigen frequencies. The SEM micrograph and the amplitude and phase response of the flat membrane are shown in Figure (to be included in the SI). The flat membrane exhibits reflection $R$ and absorption $A$, functioning as a partially reflective mirror, where radiation pressure dominates the optical force. Likewise, we also normalize the simulated force $F_z$ by the radiation pressure force given by\cite{swartzlander2017radiation}
%\begin{equation}
%    F_{RP} = \frac{P}{c}[2R+A],
%\end{equation}
%where $R$ and $A$ are the reflection and absorption of the unstructured (flat) membrane, $P$ is the incidence power and $c$ is the speed of light. 

%Figure \ref{fig:3} presents the comparison: the blue line represents the simulated normalized force  $F_z/F_{RP}$ (left y-axis), while the red square dots denote the normalized measured amplitude $\hat{Z}/\hat{Z}_{mirror}$ (right y-axis). The fabrication uncertainty of $\pm$ 10 $nm$ is indicated by the error bars, while the blue shaded region accounts for this uncertainty in terms of separation $S$ in the simulated data. The results confirm that the sign of the optical force can be controlled by tuning the structural parameters $R$ and $S$. Moreover, the measured and simulated data exhibit good agreement within the given fabrication uncertainty.

% 8 Positon dependence of force (Fig 4)  ===============================================
%\paragraph{Position dependence of the force}
\begin{figure}
\centering
\includegraphics[width=0.5\textwidth]{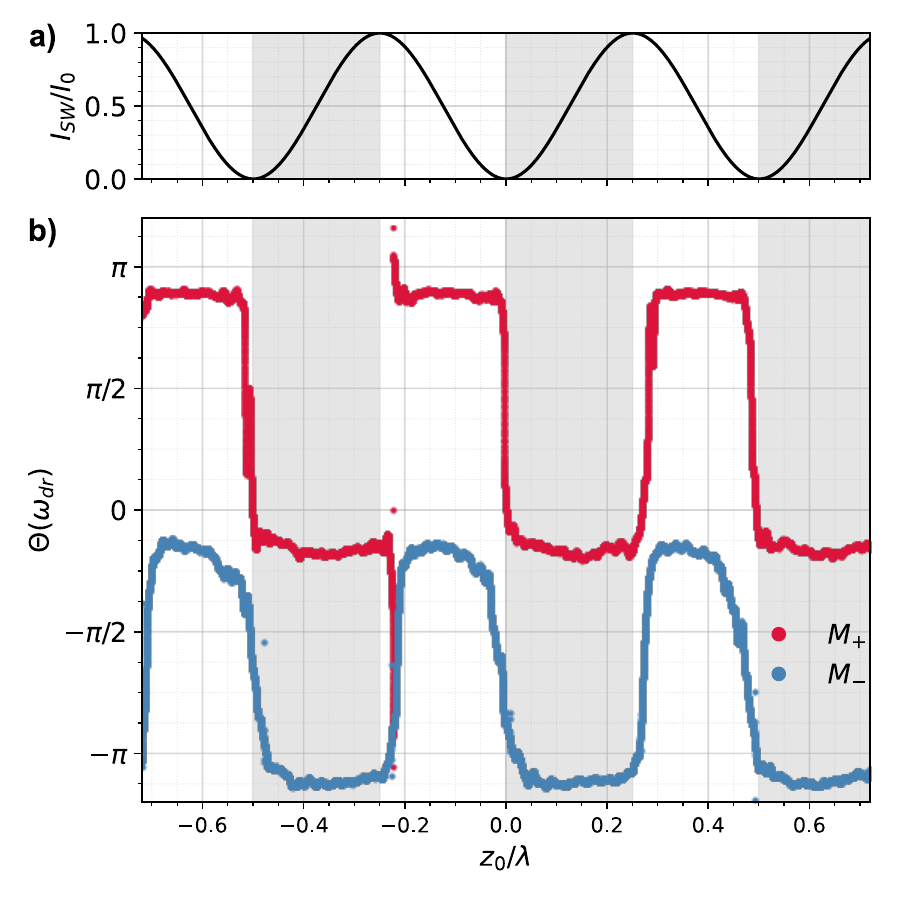}
\caption{\textbf{Position dependence of optical forces $F_\pm$.} \textbf{a}, Theoretical, normalized intensity distribution $I_\text{SW}/I_0$ along the optical axis $z$ exhibiting intensity minima and maxima. Gray (white) shaded areas correspond to a positive  (negative) intensity slope $dI/dz > 0$ ($dI/dz <0$). \textbf{b,} Phase response $\Theta(\omega_\text{dr})$ of membranes $M_+$ (red) and $M_-$ (blue). Membrane $M_+$ exhibits in-phase motion, acting as a high-field seeker whereas $M_-$ shows out-of-phase motion, behaving as a low-field seeker. Both membranes experience a phase jump of $|\Theta(\omega_{\text{dr}})|=\pi$ at points where $dI/dz =0$. The driving frequency for both membranes is $\omega_{\text{dr}}= 2\pi \times\SI{85}{kHz}$ which is below their respective resonance frequencies $\Omega_\pm$.}
\label{fig:4}
\end{figure}

As noted above, the sign of the optical force depends on both the membrane’s high- or low-field seeking behavior ($\phi_s$) and its position within the standing wave, determined by the intensity gradient ($\phi_0$). To demonstrate the position dependence of the relative phase $\phi$ between optical force $F(t)$ and reference signal, we sweep the standing wave relative to the membrane’s position $z_0$ across several wavelengths by tuning the phase $\phi_0$, while simultaneously driving the membrane's motion at $\omega_{\text{dr}}/(2\pi)=\SI{85}{\kilo\hertz}$. %applying a ramp voltage to a piezo mirror (see Fig.? in SI). This free-driving method modulated the phase difference $\Delta \phi$ between the counter-propagating pump beams from $-3\pi$ to $3\pi$, effectively translating the standing wave by three full periods across the membrane. The structures were driven at $\omega_{\text{dr}} = 85$ kHz, and 
For this analysis, we deploy $M_+$ and $M_-$ (see Tab.~\ref{tab:1}).  
%We record the phase $\Theta(\omega_{\text{dr}})$ via lock-in detection and the relative light phase$\phi_0$ by monitoring the intensity of the two counterpropagating interfereing beams on a photodiode.
%selected metasurfaces $M_+$ ($R = 345$ nm, $S = 530$ nm) and $M_-$ ($R = 485$ nm, $S = 430$ nm). 
Figure \ref{fig:4}a shows the theoretical intensity $I_\text{SW}/I_0$  along the standing wave. %as a function of $\Delta \phi$. 
The measured lock-in phase $\Theta(\omega_{\text{dr}})$ for $M_+$ (red) and $M_-$ (blue) is shown in Fig. \ref{fig:4}b, displaying a constant phase difference of $\pi$ between $M_+$ and $M_-$. For a negative intensity slope ($\phi_0 = -\pi/2$), %When sweeping from maxima to minima (negative slope), 
$M_+$ exhibits a phase shift of $\Theta(\omega_{\text{dr}})\approx \pi$, while $M_-$ remains in phase at $\Theta(\omega_{\text{dr}}) \approx 0$. The phase shifts reverses sweeping  to a positive intensity gradient  ($\phi_0 = +\pi/2$), further confirming the opposite signs of optical forces in these two metasurfaces.

% 10 OUTLOOK AND CONCLUSION ===========================================================
\paragraph{\textbf{Conclusions}}
%\paragraph{Summary and Outlook}
In summary, we demonstrate deterministic control over both the amplitude and direction of optical forces on suspended high-refractive-index metasurfaces via advanced mode engineering. Most notably, we unveil the counterintuitive phenomenon of low-field-seeking behavior, its attraction to intensity minima, enabled by the interplay of Mie multipolar modes. Beyond its fundamental significance, this approach establishes a powerful and versatile platform for optical force engineering, potentially relevant to light sails and optomechanics.

%%%%%%%%%%%%%%%%%%%%%%%%%%%%%%%%%%%%%%%%%%%%%%%%%
\textbf{Acknowledgements:} This research was supported by the European Research Council (ERC) through grant Q-Xtreme ERC 2020-SyG (grant agreement number 951234). We acknowledge valuable discussions with the Q-Xtreme synergy consortium. \\
\vspace{10pt}

\textbf{Author contributions} - 
A.A. performed the numerical simulations, fabricated the device, performed the measurements and analyzed the data. A.A. and B.M. designed and implemented the optical setup. N.M developed the theoretical expressions. N.M. and R.Q. conceptualized the experiments. All authors discussed the results and contributed to writing the manuscript.\\

%\textbf{Competing Interests} - 
%The authors declare no competing interests.
%%%%%%%%%%%%%%%%%%%%%%%%%%%%%%%%%%%%%%%%%%%%%%%%%%%%%%%%%%%%%%%%

%\vspace*{25cm}

%\textbf{Data availability} - 
%Source data for Figs. 2, 3,and 4 are available via the ETH Zürich Research Collection at TBA.
%%%%%%%%%%%%%%%%%%%%%%%%%%%%%%%%%%%%%%%%%%%%%%%%%

%\bibliography{references}

\pagebreak
\section*{Methods}
\subsection{Fabrication of metasurfaces}
To fabricate the metasurfaces, we employ top-down electron beam lithography. Commercially available free standing crystalline (100) silicon membranes from Norcada Inc. serve as material substrate. The membrane sample is spin coated with the AR-P 6200.04 positive photo-resist with a thickness of $\SI{230}{\nano \meter}$ followed by baking for 1 minute at 150$^\circ$C. Afterwards, electron beam exposure is carried out followed by 90 seconds in the AR 600-546 developer at room temperature. We etch the silicon membrane using HBr chemistry with an inductively coupled plasma etcher. Finally,the photo-resist is stripped off with an oxygen plasma etcher. The patterned metasurface of area $\SI{20}{\micro \meter} \times \SI{20}{\micro \meter}$ is placed in the center of a circular membrane of diameter $D = \SI{36}{\micro\meter}$ which is connected by undulated tethers to the substrate frame.

\subsection{Simulations}
The optical forces are numerically simulated using the RF module of the commercially available solver COMSOL Multiphysics. We employ a scattered field formulation, defining a meta-atom  composed of silicon discs with cross beams suspended in air, with periodic boundary conditions applied in the $x$ and $y$ directions. Perfectly matched layers are implemented along $\pm z$ to minimize boundary reflections.

\end{document}